HARD-SPHERE FLUID

**Virial equation-of-state for the hard-sphere fluid**

Leslie V Woodcock

Department of Chemical and Biomolecular Engineering
National University of Singapore, Singapore 117576
e-mail chewlv@nus.edu.sg

**Recent values for virial coefficients up to $B_{12}$, when expressed in powers of density relative to maximum close packing ($\rho_0$), lead to a closed equation-of-state for the equilibrium fluid. The series obtained converges for all densities up to $\rho_0$ where it diverges to a negative pole. Comparisons with MD data indicate the virial pressure begins to deviate from the metastable fluid at the fluid freezing density ($\rho_f$).**

Since van der Waals first modified the ideal gas equation to obtain his "second-virial" equation–of-state for the dilute gas of hard spheres (HS), there have been a great many predictions of the HS equation-of-state. Theoretical equations, whilst yielding some physical insight into the underlying physics, have all proven to be inaccurate when tested against computer experimental data. An essential starting point for the ultimate equation-of-state of the HS gas is the known virial series. The virial expansion in powers of density for the fluid equation of state of is

$$Z = 1 + \sum_n b_n \rho^{(n-1)} \quad (1)$$

Where $Z = pV/Nk_BT$ : $k_B$ is Boltzmann's constant, $b_n$ are the coefficients in dimensions of sphere diameter $\sigma^{3(n-1)}$, and $\rho$ is the number density (N/V). Exact third and fourth virial coefficients were derived analytically by Boltzmann, higher coefficients have now been computed numerically up to the eleventh [1-7]. $B_{12}$ has been obtained from MD data [6,7] All the computational details are given in the various references [1-7]

Numerical values of the higher virial coefficients are often presented as a dimensionless ratio of $b_n/b_2^{(n-1)}$ where $b_2$ is the second virial coefficient determined by van der Waals to be $b_2 = 2/3 \pi \sigma^3$. The latest recommended values for all the known virial coefficients from Clisby and McCoy [7] are in Table I. Also given in Table I are the values of the known virial coefficients, when expressed in powers of the density relative to crystal close packing $\rho_0 \sigma^3 = 2^{½}$.

$$Z = 1 + \sum_n B_n (\rho/\rho_0)^{(n-1)} \quad (2)$$

Before the coefficients $B_8$ to $B_{11}$ were known, it appeared that a simple closed-form equation-of-state might be possible on the observation that the virial coefficients in equation (2) were approaching 9, the dimension squared [8]. Computed values $B_8$-$B_{11}$ ,

however, show that this is not the case. Nevertheless, if in all dimensions, this expansion peaks very close to $D^2$, which suggests that the series does indeed reflect the value of $\rho_0$.

On revisiting the expansion in powers of density relative to the maximum, this form therefore, could contain the clue to the ultimate destiny of the higher virial coefficients.

TABLE I    Virial coefficients of the hard-sphere fluid

| n | $b_n/b_2^{(n-1)}$ [7] | $B_n$ |
|---|---|---|
| 2 | 1.00000000 | 2.961921 |
| 3 | 0.62500000 | 5.483111 |
| 4 | 0.28694950 | 7.456345 |
| 5 | 0.11025210 | 8.485568 |
| 6 | 0.03888198 | 8.863719 |
| 7 | 0.01302354 | 8.793670 |
| 8 | 0.00418320 | 8.366104 |
| 9 | 0.00130940 | 7.756405 |
| 10 | 0.00040350 | 7.079543 |
| 11 | 0.00012300 | 6.392052 |
| 12 | 0.00003700 | 5.695219 |

If we examine the trend in the virial coefficients in powers of density relative to crystal close packing as in equation 2; (figure 2) the virial coefficients are beyond $B_8$ are decreasing linearly according to

$$B_n = C - An \quad (n \geq 9) \tag{3}$$

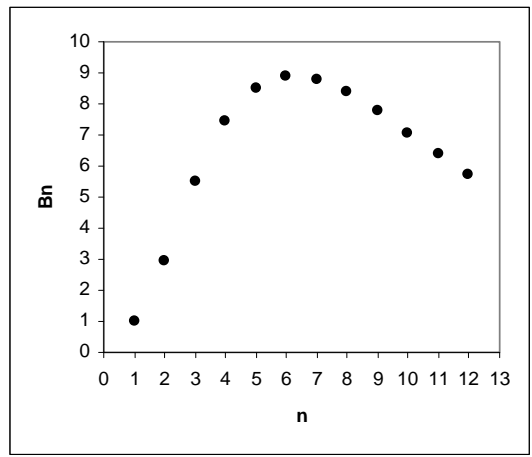

FIG.1 Virial coefficients of the hard-sphere fluid in the expansion in powers of density relative to crystal close-packing: values from Clisby and McCoy, reference.[7] $B_2$- $B_4$ are analytic, $B_5$-$B_{11}$ are numerical and $B_{12}$ was determined previously by others from MD data.

with an EXCEL trendline regression $R^2$ of 0.9999. Since there is no imaginable physical reason for any deviations from this trend, we can hypothesize that all the higher virial coefficients greater than $B_8$ can be obtained from equation (3).

If this hypothesis were to be confirmed, an essentially exact, albeit numerical, equation-of-state for the hard-sphere fluid, up to the density of the first phase transition, can be derived. If we let m denote the highest virial coefficient that is not given by equation (3), the analytic closed-form expression, for the summation of all powers of ρ greater than m, is obtained. When $B_n$ are given by equation (3), the terms in the infinite summation are a standard integral that is solvable to give the simple closed formula

$$Z = 1 + \sum_{n=2}^{m} B_n (\rho^*)^{(n-1)} + (\rho^*)^m \left[ \frac{(C-mA)}{(1-\rho^*)} - \frac{A}{(1-\rho^*)^2} \right] \quad (4)$$

where $\rho^* = \rho/\rho_0$ ; the EXCEL trendline values for the two constants are C =13.8979 and A = 0.68219.

Equation (4), with these constants, predicts that the virial coefficients eventually go negative and stay negative, the first negative coefficient being $B_{21}$ This scenario for hard spheres is consistent with the known behavior of virial coefficients for simpler hard objects, such as parallel cubes. It is also consistent with the trend of the virial coefficients in higher dimensions.

In order to test the accuracy of equation (4) molecular dynamics (MD) computations of the hard-sphere fluid pressure equation-of-state have been performed in the density region from $\rho\sigma^3$ = 0.95 to 1.025. The fluid freezing transition occurs just below 0.95 and the highest density that the extrapolated metastable branch can be accessed accurately is around 1.025 i.e. close to the crystal melting density. The MD results are compared with the pressures obtained from the virial equation of state (equation 4) in TABLE II., and shown graphically in Fig. 2.

TABLE II. Equation-of-state data for the hard-sphere fluid in the high-density metastable fluid region, compared with MD data obtained for a system of 64000 spheres : estimated uncertainty in MD data is $\pm$ 0.0002

| $\rho\sigma^3$ | 0.950 | 0.975 | 1.000 | 1.025 |
|---|---|---|---|---|
| MD | 12.79237 | 13.98365 | 15.34992 | 16.88325 |
| Eq. 4 | 12.79254 | 13.97910 | 15.30325 | 16.78372 |

# HARD-SPHERE FLUID

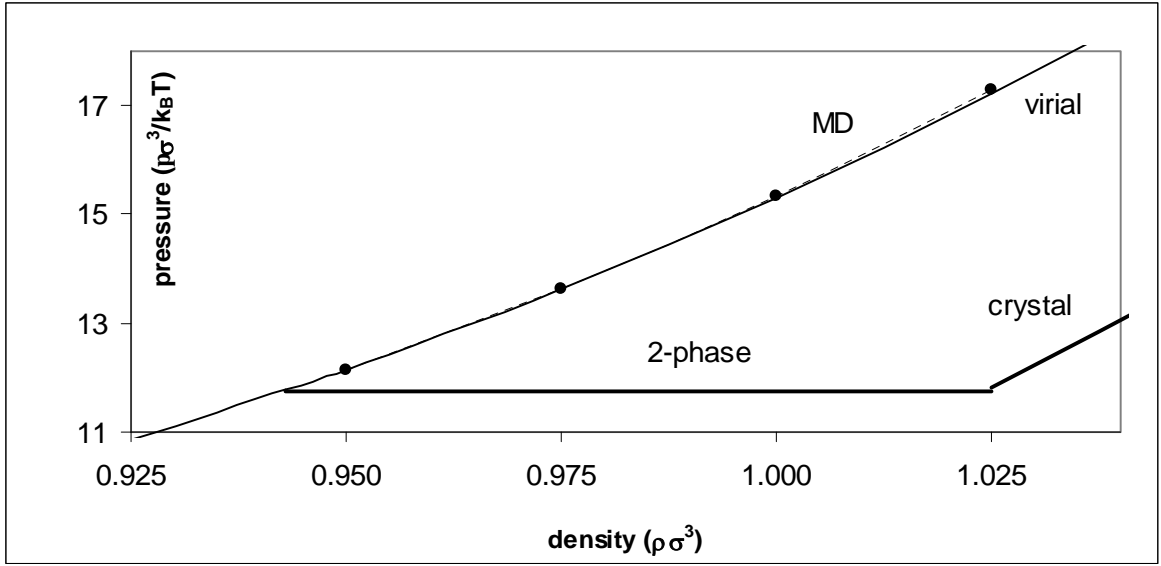

FIG.2  Pressure of the hard-sphere fluid in the vicinity of the freezing transition and the two-phase region; the virial equation (4) begins to deviate from the MD dense fluid pressure data (solid points) at the equilibrium fluid freezing density ($\rho_f$)

There is presently still some uncertainty regarding the thermodynamic status of the metastable region of the hard-sphere fluid beyond its freezing transition up to densities of random close packing (RCP). We have recently found that a well-defined RCP state can be accessed both by irreversible isothermal compactions, and also by a reversible path based upon the simple-cubic single-occupancy model. Whilst further, more accurate, numerical MD results are needed, it appears that that these metastable fluid extrapolations, and the hard-sphere gas phase from zero density, do not belong to one and the same equation-of-state that is continuous in all its derivatives.

———————————————